\documentclass[12pt,preprint]{aastex}
\usepackage{natbib}
\usepackage{epsfig}
\usepackage{graphicx}
\begin{document}

\title{Ly$\alpha$ Emission from High Redshift Sources in COSMOS}

\author{ Ryan P. Mallery\altaffilmark{1},  Bahram Mobasher\altaffilmark{1}, Peter Capak\altaffilmark{2}, Yuko Kakazu\altaffilmark{2}, Dan Masters\altaffilmark{2}, Olivier Ilbert\altaffilmark{3}, 
Shoubaneh Hemmati\altaffilmark{1}, Claudia Scarlata\altaffilmark{4}, Mara Salvato\altaffilmark{5}, Henry McCracken\altaffilmark{6}, Olivier LeFevre\altaffilmark{3}, Nick Scoville\altaffilmark{2}}
\altaffiltext{1}{Department of Physics and Astronomy, University of California, Riverside,  900 University Ave, Riverside CA 92507}
\altaffiltext{2}{California Institute of Technology, 1200 East California Boulevard, Pasadena, CA 91125}
\altaffiltext{3}{Laboratoire dÕAstrophysique de Marseille, CNRS-UniversitŽ de Provence, 38 rue FrŽdŽric Joliot-Curie, 13388 Marseille Cedex 13,  France}
\altaffiltext{4}{School of Physics and Astronomy, University of Minnesota, 116 Church St, Minneapolis, MN 55455}
\altaffiltext{5}{Max-Planck-Institut fŸr Astronomie Kšnigstuhl 17 D-69117 Heidelberg, Germany}
\altaffiltext{6}{Institut dÕAstrophysique de Paris, UMR7095 CNRS, Universite Pierre et Marie Curie, 98 bis Boulevard Arago, 75014 Paris, France}

\begin{abstract}
We investigate spectroscopically measured Ly$\alpha$ equivalent widths and escape fractions of $244$ sources of which $95$  are Lyman Break Galaxies (LBGs)
and $106$ Lyman Alpha Emitters (LAEs) at $z\sim4.2$, $z\sim4.8$, and $z\sim5.6$ selected from intermediate and narrow-band observations.
The sources were selected from the Cosmic Evolution Survey (COSMOS),  and observed with the DEIMOS spectrograph.
We find that the distribution of equivalent widths shows no evolution with redshift for both the LBG selected sources and the intermediate/narrow-band LAEs. 
We also find that the Ly$\alpha$ escape fraction of intermediate/narrow band LAEs is on average 
higher and has a larger variation than the escape fraction of LBG  selected sources. The escape fraction does not show a dependence with redshift. 
Similar to what has been found for LAEs at low redshifts, the sources with the highest extinctions show the lowest escape fractions. The range of escape fractions increases with decreasing extinction.
This is evidence that the dust extinction is the most important factor affecting the escape of Ly$\alpha$ photons, but at low extinctions other factors such as 
HI covering fraction and gas kinematics can be just as effective at inhibiting the escape of Ly$\alpha$ photons. 
\end{abstract}

\section{Introduction}
The study of the high redshift universe and the early evolution of galaxies has  primarily relied on two techniques to obtain large 
samples of high redshift galaxies, the Lyman-break technique (LBGs; Steidel et al. 1999, Ouchi et al. 2004;  Bouwens \& Illingworth 2006, and references therein) 
and narrow band surveys targeting Ly$\alpha$ emitting galaxies 
(LAEs;  Hu \& McMahon 1996; Rhoads \& Malhotra 2001; Ajiki et al. 2003; Hu et al. 2004; Taniguchi et al. 2005; Murayama et al. 2007, Gronwal et al. 2007, Ouchi et al. 2008; Hu et al. 2010, and references therein).
Studying the difference in the nature and properties of the two populations, selected by these two techniques, helps to understand early stages of galaxy formation and provides constraints on reionization. 
However, the two populations of  galaxies are found to have a degree of overlap,
with a fraction of the LBGs having Ly$\alpha$ emission (Shapley et al. 2003; Kornei et al. 2010;  Stark et al. 2010). The varying degree of overlap between the two techniques and how it changes with redshift is still an open question. 
Several authors have explored this by comparing spectral energy distributions (SED) properties of these two populations (Gawiser et al. 2007; Gronwall et al. 2007). 
Even less understood is the degree of overlap in the Ly$\alpha$ properties of the populations selected by these two techniques. \citet{kornei} recently studied the Ly$\alpha$ properties of $z\sim3$ LBGs and 
found that LBGs with strong Ly$\alpha$ emission are older, have lower SFR, and are less dusty than objects with either
weak Ly$\alpha$ emission, or the line in absorption.  They concluded that, within the LBG sample, objects 
with strong Ly$\alpha$ emission represent a later stage of galaxy evolution in which supernovae-induced outßows 
have reduced the dust covering fraction. In contrast, analysis of LAEs at $z\sim3.1$, $3.7$, and $5.7$ by Ouchi et al. (2008) have revealed that LAEs have lower extinction and/or younger ages than LBGs.

Due to the complex physics of Ly$\alpha$ radiative transfer process in galaxies, 
modeling Ly$\alpha$ emission, absorption, and escape has been investigated 
by numerous authors.  \citet{neufeld} and \citet{cf00}  modeled the Ly$\alpha$ radiative transfer and investigated
the role of a clumpy, dusty, multiphase ISM on Ly$\alpha$ escape. \citet{hansen} has expanded on these past attempts by considering the effects of several different
geometrical distributions of dust clouds, while \citet{dijkstra} and \citet{verhammeA} have incorporated
the effect of in-falling or outgoing spherical halos of neutral gas on Ly$\alpha$ escape and its profile.
In particular, the monte-carlo radiative transfer models by \citet{verhammeB} taking into account
dust, ISM kinematics, HI column densities, and gas temperature, have been able to reproduce
the Ly$\alpha$  profiles of $11$ LAEs  found in \citet{tapken}.

 Analysis of nearby Ly$\alpha$ emitting galaxies (Kunth et al. 2003;  Mas-Hesse et al. 2003; Hayes et al. 2005; Ostlin et al. 2009; Atek et al. 2009; Scarlata et al. 2010) indicates that Ly$\alpha$ emission is affected by ISM geometry, gas kinematics and
dust. However, the order of importance of each of these factors is not clearly established and could
possibly vary from object to object \citep{schaerer}. One method to ascertain the principle physical factors that affect the Ly$\alpha$ radiative transfer in galaxies, is to measure the Ly$\alpha$ escape fraction (f$_{esc}$), defined as
the ratio of the  observed Ly$\alpha$ flux to what is expected from the star formation rate (SFR) of the galaxy. 
In recent years the study of the escape fraction of Ly$\alpha$ photons in star forming galaxies at redshifts ranging from $z\sim0.1-6$ has been studied by several authors 
(Scarlata et al. 2009; Finklestein et al. 2009; Atek et al. 2009; Hayes et al. 2010; Ono et al. 2010a; Ono et al. 2010b). Each study has found a strong trend of decreasing escape fraction with increasing extinction, though
any change in the mean escape fraction of Ly$\alpha$ sources with redshift is uncertain given the difference in the methods of selecting samples of Ly$\alpha$ sources at  $z\sim0.1$, $z\sim2$, and  $z>3$.

In order to examine the varying degree of overlap between the Ly$\alpha$ properties of these two populations (LBGs and narrow band selected LAEs) and its redshift dependence,
deep spectroscopic observations are required to measure the fraction of LBGs with Ly$\alpha$ emission.
Spectroscopic follow-up for these high redshift sources has only recently been made possible  due to the technical difficulties in the spectroscopy of faint, m$_{I} >22$, high redshift sources.
Ouchi et al. 2008 obtained  Subaru/FOCAS and VLT/VIMOS spectroscopy of 84 out of 858 narrow band LAE candidates at  $z= 3.1$, $3.7$, and $5.7$. 
The Ly$\alpha$ luminosity function of these sources increases with redshift, indicating that galaxies with Ly$\alpha$ emission
are more common at higher redshifts.
\citet{hu2010} presented an atlas of $88$ $z\sim5.7$ and $30$ $z\sim6.5$ spectroscopically confirmed LAEs.   
Ouchi et al. 2010 presented spectra of LAEs at $z\sim6.6$ examining the Ly$\alpha$ line profiles, the luminosity function, clustering properties of the sources. 
Analysis of their sample in comparison with LAEs at z$\sim5.7$ indicates that the intergalactic medium (IGM) was 
not highly neutral at $z\sim6.6$ and the bulk of reionization of the universe occurred at $z>7$. \citet{stark2010} spectroscopically confirmed $199$ Ly$\alpha$ galaxies from a sample of $627$ continuum selected LBGs at $3<z<7$ and found 
that the fraction of LBGs with Ly$\alpha$ emission increases with redshift and is inversely correlated with UV luminosity. 
The likely cause of this is a decrease in dust extinction with redshift, and also a lower HI covering fraction for sources with lower UV luminosity.

In this paper we study Ly$\alpha$ emission from sources at  $4<z<6$, detected in deep spectroscopic survey of the COSMOS field \citep{scoville}.
The selected sources  consist of intermediate and narrow band LAEs at $z\sim4.2$ (IA624), $z\sim4.8$ (NB711) and  $z\sim5.7$ (NB816), B$_J$ LBGs, g$^+$ LBGs, V$_J$ LBGs, r$^+$ LBGs, i$^+$ LBGs, and sources with  photometric redshifts $z > 4$ . 
In \S2, we present the data, and the method used for source selection.
In \S3, we present our analysis of the Ly$\alpha$ emission as it relates to both redshift and our source selection. In \S4 we estimate the Ly$\alpha$ escape fraction  and perform a speculative analysis based on our estimates. 
Our conclusions are presented in \S5.
We assume  {\it H$_o$}$=70$
km s$^{-1}$ Mpc$^{-1}$, $\Omega_{m}=0.3$, and  $\Omega_{\Lambda}=0.7$. We assume AB magnitude. s

\section{Data}
	\subsection{DEIMOS Observations and Data Reduction}
		A total of $4267$ sources were targeted for spectroscopic observations with the DEIMOS multi-slit spectrograph \citep{faber} on the Keck II telescope.  Full details of the observations and data can be found in Capak et al. (in prep).
		A total of $42$ separate slit masks were observed, each with on average $102$ $1^{\prime\prime}$ slits per mask. 
		The observations were taken over a period of several semesters with 5 nights in January 2007,  4 nights in November 2008, 4 nights in November 2009, 7 nights in January 2010 and 5 nights in February 2010. 
		The observations were taken with the 830 line BK7 grating with a wavelength coverage of $\sim6000-9000$\AA. Observations of each mask were dithered by $1^{\prime\prime}$  with a total integration of  3.5 hours for each mask.
		Reductions were performed creating 1d spectra for each slit, using a variation of the standard DEIMOS spec2d reduction package in order to account for the dithered observations.  Flux calibration was  performed by first using stellar spectra 
		to measure the detector response profile for each mask. The 1d spectra were then divided by the response profile and normalized.
		For absolute flux calibration, the spectra were then integrated over Subaru filter response profiles and scaled by the error-weighted mean ratio between magnitude 
		(computed from the spectra) and Subaru photometry.  Multi-bandpass Subaru photometry were used consisting of  broad (r, i, z), narrow  (NB711, NB816) and intermediate (IB624, IB709, IB738, IB767) 
		band filters from the publicly available COSMOS optical catalog 
		(see Capak et al. 2007)\footnote{http://irsa.ipac.caltech.edu/data/COSMOS/tables/photometry/
		This catalog includes the photometry in all the 25 optical/NIR broad-, intermediate- and narrow-bands filters, from â"u" to â"Ks" . The photometry is computed at the position of the i*-band image, using Sextractor (Bertin \& Arnouts 1996) in dual mode. 
		The catalog supersedes Capak et al. (2007), with improved source detection and photometry extracted in 3$^{\prime\prime}$ apertures.} 
		 The flux calibration  procedure used, removes any slit-loss as the spectroscopy is scaled directly to the photometry.  

	\subsection{Source Selection}
	  A total of $1453$ of the observed sources were selected to be at $z > 3.8$. After examination of their spectra, and removal of stellar sources and low-z interlopers the number of possible $z > 3.8$ sources is $644$. 
	  The goal of the Keck program was to select as complete a sample at $z>4$ as possible, for objects brighter than $z^+<25$ and more massive than $10^{10.5}M_\sun$ (Capak et al. in prep).  
	 To achieve this goal, a set of continuum selected objects brighter than $z^+<25$ or IRAC [$4.5\mu$m]$<23.5$ were selected to satisfy the above magnitude and mass limits respectively.  
	 From this flux limited sample  $B_J$, $g^+$, $V_J$, $r^+$, $i^+$, and $z^+$ LBGs
	  were selected using known  criteria (Ouchi et al. 2004, Capak et al. 2004, 2011b, Iwata et al. 2003, and Hildebrandt et al. 2009).  Objects with a probability greater than 50\% of 
	  being at $z>4$, based on the Ilbert et al. (2010) photo-z catalog, were also 
	  included if they met the flux limit.  Finally, to avoid any biases against heavily dust obscured objects (e.g. Capak et al. 2008, 2011a), sources meeting the LBG or photo-z criteria and also detected by {\it Chandra}, {\it Spitzer} MIPS (24$\mu$), 
	  AzTEC (1.1mm), Mambo (1.24mm), BoloCam (1.1mm) or the VLA (20cm) were also included in the sample even if they were fainter than the flux limit.  
	  
	  In addition, Ly$\alpha$ emitters were selected using the IA624, NB711, and NB816 bands 
	  following previous studies (Scarlata et al. in prep, Shioya et al. 2009, Murayama et al. 2007), with the modification that a fixed color cut was used to the faintest magnitudes as done in Hu et al. (2010) instead of a noise adjusted cut. 
	  To the NB711 sources selected by the Shioya et al. (2009) criteria, sources were also added with 0.3 magnitude excess between the NB711 and the interpolated $r^+$ $i^+$ photometry, and also sources with a 0.3 magnitude excess between the NB711 
	  and interpolated IA707 and IA738 magnitudes in order to  add sources possibly having lower Ly$\alpha$ equivalent widths than the Shioya et al. (2009) selection criteria. 
	   To the NB816 sources selected by the  Murayama et al. (2007) criteria, sources were also added with 0.3 magnitude excess between the NB816 and the interpolated $i^+$ $z^+$  photometry, and also sources with a 0.3 magnitude excess between the NB816 
	  and interpolated IA707 and IA738 magnitudes in order to  add sources possibly having lower Ly$\alpha$ equivalent widths than the Murayama et al. (2007) criteria.
	 	
A total of  895 LBG sources were targeted for spectroscopy. Removal of low-z contaminants and stars leaves 380  $z > 3.8$  LBG candidates. The Suprime-Cam z$^{\prime}$ magnitudes 
of the targeted LBGs range from 22.7 to 25 AB, with a mean of 24.8 AB.
left panel of figure 1 (we refer to figure 1 again in section 2.4) shows the z$^{\prime}$ magnitude distribution of all the LBGs with spectroscopically measured redshifts.
In addition to the LBGs, 83 IA624 LAEs at $z\sim4.2$, 83 NB711 sources at $z\sim4.96$, and 98 NB816 sources at  $z\sim5.7$ were targeted for spectroscopy. 
After removal of stellar sources and low-z contaminants, the distribution of LAEs becomes
 26 at $z\sim4.2$ (IA624), 42 at $z\sim4.8$ (NB711) and 73 at  $z\sim5.7$ (NB816).	 
The IA624 sources have z$^{\prime}$ magnitudes range from 24.9 to 26.7 AB, with a mean of 25.8 AB. 
The NB711 sources vary in z$^{\prime}$ magnitudes  from 23.6 to 27.2  AB with a an average of 24.9 AB,
and the magnitudes of the  NB816 sources vary from 24.1 to 27.2 AB with a mean of 25.6 AB. The right panel of figure 1 shows the z$^{\prime}$ magnitude distribution of the IA624, NB711 and NB816 sources with spectroscopically measured redshifts. 

   For the entire sample of $644$ high redshift candidates, $244$ have $3\sigma$ detections of Ly$\alpha$, with 86 having rest-frame Ly$\alpha$ equivalent widths (EW$_{Ly\alpha,0}) > 25$\AA. Table 1 lists the number of high 
		redshift candidates, the sub-set with $3\sigma$ Ly$\alpha$ detections, and  the number with EW$_{Ly\alpha,0}>25$\AA~for each source type. 
		Of the $380$ $B_J$,$g^+$,$V_J$,$r^+$,and $i^+$ LBGs observed, $95/380$ (32/380) have $3\sigma$ detections of Ly$\alpha$ (EW$_{Ly\alpha,0} > 25$\AA):  10/49 (3/49) $B_J$,  21/158 (3/158) $g^+$,  
		39/101 (16/101) $V_J$, 23/56 (9/56) $r^+$ and 2/16 (1/16) $i^+$.  The low number of $i^+$ LBG sources with Ly$\alpha$ is likely due to low 
		number statistics and the limit of our survey (  z$^+<25$), which selects only the bright sources ( M$_{UV}<-22$) at $z\sim6$ and the color selection criteria which selected mostly stars (98/114). 
		We also find that 21/26 (9/26) of the IA624,  
		25/42 (9/42) of the NB711, and 60/73 (26/73) of the NB816
		selected sources have $3\sigma$ detections of Ly$\alpha$ (EW$_{Ly\alpha,0} > 25$\AA). \\

	\subsection{Redshift, AGN, and Ly$\alpha$ Identification}
	Of the 644 high redshift candidates observed, 372 have high quality/reliable redshifts at $z>3.8$. Each spectrum was examined by eye in IDL using {\bf SpecPro} \citep{masters} by at least two people, and often by three (RM, DM, \& CP). 
	Spectra with Ly$\alpha$ were easily identified by its asymmetric emission line shape (see figure 2). 
	Spectra with only low S/N absorption features required 
	several features before being confirmed. This included spectroscopic redshifts consistent with the photometric spectral energy distribution (SED), 
	and agreement between independent estimates of the spectroscopic redshift.

	The contamination of the high redshift sources by AGN is not well known. At the flux limits for the XMM survey of COSMOS \citep{cappelluti},  we expect detections of only the high redshift sources with L$_{X}>10^{45}$ ergs/s.
	This is over three orders of magnitude higher than the standard AGN X-ray detection limit  L$_{X}>10^{42}$ ergs/s.  No sources are individually detected by XMM. 
	 One high redshift source ($\alpha=150.35980~\delta=2.0737081$) is detected in the X-ray  by {\it Chandra} in the C-COSMOS survey (Elvis et al. 2009), though unlike the XMM survey of COSMOS the {\it Chandra} survey is not uniform over the entire field.
	Two of the LBG sources are point sources in ACS ($\alpha=149.87082~\delta=1.8827920$, and $\alpha=150.13036 ~\delta=2.4660110$  taken from Ikeda et al.  2011), but show no signs of AGN in their spectra, nor have x-ray  detections.

	Spectroscopic identification of AGN via NeV$\lambda$$1238$ emission or other broad emission lines ([CIV]$\lambda$$1550$ and CIII$\lambda$$1908$) is largely dependent on their redshifts.
	CIII$\lambda$$1908$  is redder than the wavelength cutoff
	for sources at $z>4.2$, and [CIV]$\lambda1550$ for sources at $z>5.4$. A total of 15/644 sources show possible signs of AGN in their spectra, with 6 of these also having Ly$\alpha$ detections. 
	Including the {\it Chandra} detection, this gives a lower limit of $2.9$\% (7/244) for AGN contamination in our Ly$\alpha$ sample. 
	AGN contamination in sources with Ly$\alpha$ emission have been reported at $43$\% at $z\sim0.1$ \citep{finkelstein}, 3\%$-$7\% at z = 2.1 (Guaita et al. 2010), 5\%$-$13\% at $z\sim2.25$ (Nilsson et al. 2009), 
	1\%$-$10\% at $z\sim3.1-3.7$ (Gronwall et al. 2007; Ouchi et al. 2008; Lehmer et al. 2009), $<3.2$\% ($<6.3$\%) for type-1 (type-2) AGN at $z\sim4.5$ (Zheng et al. 2010),   
	$<5$\% at $z\sim4.5$ (Malhotra et al. 2003; Wang et al. 2004), and $<1$\% at $z\sim5.7$ (Ouchi et al. 2008).

	\subsection{Selection Bias}	
		Selection bias for the sub-sample of high redshift spectroscopic sources with Ly$\alpha$ emission is expected to be low as the spectroscopic sample was selected to be complete at $z>4$ for objects brighter than $z^+<25$ and 
		more massive than $10^{10.5}M_\sun$.  Ly$\alpha$ emission is detected to a redshift dependent flux limit of $\sim5e-18$~ergs/s/cm$^2$.
	        Figure 1 shows the redshift plotted versus the  $z^+$ (AB) magnitude for all high redshift candidates with reliable spectroscopic redshift.
		Down to the limits of our survey (  $z^+<25$) 
	there appears to be no bias between sources with Ly$\alpha$ detections and those without, for both LBG and intermediate/narrow band selected sources. For the high redshift candidates observed using other selection 
	criteria, the number statistics are too low for a meaningful comparison.
	
		The amount of overlap between the LBGs and the intermediate/narrow-band selected LAEs is not fully known. In principle we can check which (if any) of the intermediate/narrow-band LAEs satisfy the color conditions used to select the LBGs.
	However, many of the intermediate/narrow-band LAEs are too faint and  not detected in many of the various bands used to create the LBG source list. As figure 1 shows, 
	most of the intermediate/narrow-band LAEs are fainter than the $z^+<25$ criteria used to create the LBG source list. Relaxing this criteria for the intermediate/narrow-band LAEs, 
	we can check the LBG color criteria for the intermediate/narrow-band LAEs that have the appropriate detections in the broad band photometry. For the 21 IA624 Ly$\alpha$ sources, 11 would be considered either $B_J$, $V_J$, or $g^+$ LBGs, 
	2 do not match any of the LBG criteria, and 8 are not detected in the enough bands to say one way or the other. For the 25 NB711 Ly$\alpha$ sources, 17 would be considered either $B_J$, $V_J$, $g^+$ or $r^+$ LBGs, 
	2 do not match any of the LBG criteria, and 6 are not detected in a sufficient number of bands to say one way or the other.  For the 60 NB816 Ly$\alpha$ sources, 9 would be considered either  $V_J$, or $r^+$ LBGs, 
	23 do not match any of the LBG criteria, and 28 are not detected in the enough bands to anything definite. 
	
        \subsection{Fraction of LBGs with Ly$\alpha$}
        	The fraction of LBG sources with Ly$\alpha$ emission has recently become a potentially important ratio, as a decrease in this fraction at $z > 6$ may be indicative of an increase of the neutral fraction of gas in the 
	intergalactic medium (Furlanetto, Zaldarriaga \& Hernquist 2006,
	 Mesinger \& Furlanetto 2008, Dayal, Maselli \& Ferrara 2011). Currently, there has been some debate over whether such a trend has been detected.  
	 The luminosity functions of narrow-band LAEs studied by Kashikawa et al. (2006) and Ota et al. (2008) have shown a
	 decline between $z = 5.7$ \& $z = 7.0$ indicating that the  IGM  becomes increasingly neutral above $z > 6$, while those of Tilvi et al. (2010) and Krug et al. (2011) for narrow-band LAEs at $z=7.7$ are consistent with no evolution.

	 Several authors (Curtis-Lake et al. 2011, Stark et al. 2010, 2011, Schenker et al. 2011) have measured the fraction of LBG selected sources with spectroscopically detected Ly$\alpha$ emission at $z > 4$.
	At  $z\sim7$ Ono et al. (2011), Pentericci et al. (2011) and  Schenker et al. (2011) all find that the fraction decreases from  $z\sim6$ to $z\sim7$.
	Currently, there is a factor of 2 discrepancy between the fraction of luminous dropout sources with EW$_{Ly\alpha,0} > 25$\AA~at $z\sim6$ (Curtis-Lake et al. 2011, Stark et al. 2010).
	Figure 3 shows the fraction of LBGs with EW$_{Ly\alpha,0} > 25$\AA~and $-20.25 <$~M$_{UV} < -21.75$.  A completeness correction was made by adding simulated EW$=25$\AA~lines into the spectra (by RM), and having 
	another author (SH) blindly search and measure the simulated lines. The mean completeness for the LBGs with EW$_{Ly\alpha,0} > 25$\AA~and m$_{continuum} < 26$ (AB)  is  95\%.
	In figure 3, the $B_J$ and $g^+$ LBGs are plotted together as a lower limit, since the   the color selection criteria can select sources with redshifts below the minimum redshift that Ly$\alpha$ can be measured  for the spectroscopic  setup used.
	For the $B_J$ and $g^+$ LBGs at $<$z$>\sim4.2$, we calculate lower limit of  $5$\%; for the $V_J$ LBGs at $<$z$>\sim4.6$, we get a fraction of $18\pm12$\%; and for the $r^+$ LBGs at $<$z$>\sim5$, a fraction of $15\pm16$\%. 
	These values agree, within the errors, with the fraction of LBGs with Ly$\alpha$ reported by Stark et al. (2010, 2011) and Schenker et al. (2011). Our estimates are below those reported by 
	Curtis-Lake et al. (2011) and Stark et al. (2010, 2011) at $z\sim6$, and do not support evolution in the fraction of LBGs with Ly$\alpha$ over  the redshift range $3.8 < z < 5.5$.

	\subsection{Ly$\alpha$ Measurements}
	 A detailed procedure is used to measure the flux, equivalent width (EW), peak wavelength and full width half maximum (FWHM) of the Ly$\alpha$ emission line in the spectra. 
	 Among the issues to overcome with the data concerning these measurements, is the faintness of the continuum, its low S/N$\lesssim1$, and the varying shape of the Ly$\alpha$ 
	 feature which does not necessarily ascribe to one consistent  mathematical form from one source to the next. Variations in the continuum particularly effect the accuracy of our 
	 equivalent width measurements. In order to better elucidate our techniques, we first describe the particular method for ascertaining each measurement, and then describe the overall procedure. 
	 For several of the DEIMOS-COSMOS sources, the Ly$\alpha$ emission is double-peaked, with the wavelengths between the two peaks containing only detections of photons at the level of the continuum. These features are not [OII] as 
	 the long wavelength features shows a strong asymmetry, and the wavelength separation is always at least 5\AA~greater than would be expected if the features were [OII] doublets.
		For these cases the flux, equivalent width, peak wavelength, and FWHM are measured simultaneously for both peaks. Estimates of these quantities  are made both from a skewed Gaussian fit to the data, and from numerical methods. 
		A model for the skewed Gaussian is given in equation 1, with example spectra shown in figure 2. 
		The fit returns values for the flux normalization (A), the first moment of a standard Gaussian ($\lambda_0 = x + \omega\delta \sqrt{2/\pi}$ ), the second moment of a standard Gaussian ($\sigma = \omega \sqrt{1 - 2\delta^2/\pi}$), 
		the value of the skew ({\it s}), and the value of the continuum (c), where  $\delta = s/\sqrt{1+s^2}$.
		In figure 2, the skewed Gaussian fit to the Ly$\alpha$ line is shown in red, with the region used for numerical integration  of the flux and equivalent width shown in blue.
		The flux, equivalent width, peak and fwhm of the Gaussian and their associated errors are derived by  fitting equation 1 to the data.

		\begin{eqnarray} 
			 flux =  & A * e^{-0.5 * (( \lambda - x)/\omega)^{2}} (\int_{-\infty}^{s( \lambda - x)/\omega} \! exp(-t^{2}/2) \, \mathrm{d}t) +  c	
			      	\end{eqnarray} 
		
	         	To determine the peak wavelength of the Ly$\alpha$ emission, we first calculate the derivative of each spectrum numerically.
		The peak is then taken to be the wavelength of the emission feature where this derivative is zero. 
		The flux is then measured by numerical integration of the data, using Simpson's rule, where the continuum of the Gaussian fit is subtracted from the spectrum. The wavelength bounds for the numerical integration are determined
			by first nearest neighbor smoothing the spectrum. The bounds used for the numerical integration are then the first pixels in the smoothed spectrum
			nearest to the peak that fall below the  continuum of the Gaussian.  The region used for numerical integration is illustrated in figure 2.
			Using these bounds, the unsmoothed spectrum  minus the continuum is numerically integrated. 
			In order to estimate the error, the numerical flux integration is repeated $500$ times, each time the spectrum is varied randomly by the error of each pixel.
			The error of the numerically integrated flux is the standard deviation of $500$ the iterations.  Increasing the number of iterations was found to have a negligible effect on the determined errors of the flux, EW, and FWHM.
			
		   The EWs  are numerically integrated via Simpson's rule with the same boundaries as the flux, 
		        and the same continuum value from the Gaussian fit.  We impose the criteria that the continuum determined by the Gaussian be positive and only determine the EW for these cases.
		        	The spectra was used to determine the continuum instead of the  broadband photometry in order to limit any biases that may be introduced due an assumption of the UV slope.
		         The EW error is calculated in a similar fashion as the measurement of the flux errors.
			However, the distribution of the EWs tend to be skewed to lower values due to the faintness and 
			low S/N detection of the continuum for most of the sources. Therefore,  the standard deviation is a bad representation of the error. 
			Instead, the  $15.9$\% and $84.1$\% percentile values of the distributions are reported.  The EWs are then converted to rest-frame EWs by dividing by ($1+z$).	
			In figure 4, we compare the EWs measured using the continuum from the spectra, versus EWs measured using continuum fluxes derived from the photometry. The continuum flux at 1215\AA is derived from the photometry
			by  quadratic interpolation of the photometry for each source from each band (listed in \S2.1) with at least a $5\sigma$ detection. Only 104 sources have photometric detections to the red and blue (or at the wavelength) of the Ly$\alpha$ line
			to constrain the continuum flux at Ly$\alpha$ from the photometry.  The EWs are consistent within the errors for 75\% of the sources, and only 4\% have greater than a $2\sigma$ deviation.

			The  FWHM is measured from the spectra by first fitting b-splines to  the blue side of the peak pixel, and another to the red side of the peak pixel.
			Each spline is mirrored and the FWHM is then measured for each. The FWHM is taken as the average of the FWHM for two splines. 
			This procedure is repeated 500 times varying the spectrum by its errors as in the other numerical calculations, and  
			the error of the FWHM is taken to be the standard deviation of the 500 FWHM simulations.

		The procedure we use to incorporate each of the measurements described above also takes into account how the wavelength boundaries used for the Gaussian fit affects our measurements and errors.	
	         First, for each Ly$\alpha$ emission feature, the spectrum is  smoothed with 3-pixel boxcar and fitted with the skewed Gaussian in equation 1, using MPFIT \citep{mpfit} in IDL, 
	         without specifying the wavelength range around the emission line.
	         This fit is used to make an initial estimate of the continuum, the centroid and width of the emission feature. (NOTE: For the sources with two peaks, both features are fitted simultaneously).
		The wavelength boundaries for the numerical integration are estimated, and the skewed Gaussian is again fit to the data but only to the continuum on the red-side of the emission peak.
		Next, an iterative procedure is applied  to  compensate for any systematics that are introduced from the choice of the  continuum region
		that is used in the fit. The skewed Gaussian is fit to the data  covering a wavelength range from the short wavelength boundary used for numerical integration out to  $\lambda_{0} + 4*\sigma$. 
	         The coefficients and errors on the coefficients for the skewed Gaussian fit  are used to calculate the flux, EW, peak and FWHM of the skewed Gaussian.  
	         	As detailed above, the wavelength boundaries for the numerical integration are determined and the flux, EW, peak, and FWHM and corresponding errors are calculated.
	    	The wavelength range is increased on the long wavelength side of the centroid by  $\lambda_{0} + 4*\sigma$ + $1$pixel, and 
		a new skewed Gaussian is fitted to the data and the measurements are calculated again.
	         This is iteratively done until the   boundaries for the Gaussian fit   are equal to   $\lambda_{0}  +10\sigma$.
		This usually needs $\sim30$ iterations for each Ly$\alpha$ feature.
		The median of the flux, EW, and FWHM,  is taken as our best estimate, and except for the EWs, the standard deviation for each is added in quadrature to the error estimates from the individual iterations to obtain our final error estimates. 
		 For the EW errors, every equivalent width calculation made for every iteration is  placed into a single distribution and the $15.9$\% and $84.1$\% percentile values are taken as the error on the 
		 numerically integrated equivalent widths.  Table 2 shows the numerically estimated values for sources with a single Ly$\alpha$ peak and Table 3 shows the values for the sources with both a blue and redshifted Ly$\alpha$ peak.

\section{Equivalent Width and Redshift Distribution}
The  redshift distribution of the Ly$\alpha$ sources is shown in figure 5 and the distribution of  EW$_{Ly\alpha,0}$ is plotted in figure 6. 
These are divided into three categories: the total sample, the intermediate/narrow band LAEs, and the LBGs.  
The mean (median) EW$_{Ly\alpha,0}$ stay roughly constant with redshift but have a larger sample variance with increasing redshift for LBGs from 
$21.9 (19.6) \pm9.0$\AA~for $B_J$ LBGs, $19.5 (20.8) \pm9.9$\AA~for $g^+$ LBGs, $25.4 (21.1) \pm14.1$\AA~for $V_J$ LBGs, $25.0 (20.8) \pm19.4$\AA~for $r^+$ LBGs. 
The mean (median) EW$_{Ly\alpha,0}$ for the intermediate/narrow band LAEs show a similar trend with redshift and a larger variance with redshift, from $27.2 (25.0) \pm10.9$\AA~for IA624 LAEs and $21.9 (23.5) \pm9.5$\AA~for NB711 
selected sources to $26.6 (24.9) \pm14.1$\AA~for NB816 selected sources. A comparison between the Ly$\alpha$ properties of the intermediate/narrow-band LAEs and the LBGs at similar redshifts will be instructive.
While, unfortunately there are too few $i^+$ LBGs to compare with the NB816 selected sources, a comparison can be made between the $g^+$ LBGs and the IA624 LAEs 
as well as the $V_J$ LBGs and the NB711 sources. The $g^+$ LBGs and the intermediate band IA624  LAEs  both have the same number of sources ($21$) and the number of sources in the 
NB711 sample (24) is roughly $3/5$  the number $V_J$-dropouts  (39).   The  IA624 LAEs have a slightly higher mean and a larger distribution of EW$_{Ly\alpha,0}$ than the  $g^+$ LBGs, while the  $V_J$ LBG sample
has a larger mean EW$_{Ly\alpha,0}$ and  a larger variance than the NB711 sources. Comparing EW$_{Ly\alpha,0}$ for only the $V_J$-dropouts with  NB711 LAEs with similar magnitudes ($z^+<25$) 
though brings their median values into agreement  at 21.2\AA~and 21.0\AA~respectively.
None of the IA624  LAEs are brighter than $z^+<25$ to compare with the $g^+$ LBGs, but it is likely that the differences between the Ly$\alpha$ distributions for the LBGs and LAEs at a given redshift is due to the narrow band sample 
being fainter than the LBG sample.

The EW$_{Ly\alpha,0}$ for our entire sample are plotted versus redshift in figure 7. We find that the median EW$_{Ly\alpha,0}$ for the LBG and LAE sub-samples stay roughly constant with redshift.
At $z<3$, an increase in the distribution of EW$_{Ly\alpha,0}$ with redshift has also been reported by \citet{nilsson}. They found that the distribution of EWs for $z\sim3$ LAEs studied by \citet{gronwall} 
was higher than the distribution of EWs for their sample of LAEs at $z\sim2.25$. They speculated that the change in EW distributions with redshift is the result of increased dust content  in LAEs at lower redshifts.
An increase in Ly$\alpha$ EWs with redshift  has also been discovered in LBGs. \citet{stark2010} found in their sample of $\sim199$ LBGs with detected Ly$\alpha$ emission at $z=3-6$, that the prevalence of large EWs increases moderately with redshift. 

Several authors (Shapley et al. 2003; Stark et al. 2010) have noted an anti-correlation between UV luminosity and EW. This has been refuted by \citet{nilsson} who argued that the lack of luminous sources with high EWs
may be due to the fact that luminous sources and sources with high EWs are both rare, and that this parameter space has been poorly represented in current flux limited surveys. \citet{kornei}  found only a marginal correlation
between the EWs and UV luminosities for a large sample of LBGs  at $z\sim3$, with M$_{UV}<-20$. In the sample of LBGs studied in \citet{stark2010}, which detects sources to M$_{UV}=-18$, the authors found low-luminosity LBGs (M$_{UV}=-19$)
to show strong Ly$\alpha$ emission much more frequently than luminous systems (M$_{UV}=-21$). For our sample, no correlation is found between the EWs and UV luminosities, neither for the full sample nor for the LBG selected sources. 
This is likely to be a selection effect as our LBG selected sources are mostly bright, with  M$_{UV} <-20$.

\section{Estimating the Escape Fraction}
The simplest method to estimate the escape fraction is to measure the flux of both Ly$\alpha$ and extinction corrected H$\alpha$, assume a recombination regime (usually CASE B recombination, Osterbrock 1989), and compute the number of detected
Ly$\alpha$ photons divided by the number of expected Ly$\alpha$ photons estimated from the H$\alpha$ flux.  For the redshifts of our sources, H$\alpha$ is redshifted to the near-infrared and is currently unaccesible. We can however make
a crude estimate of the escape fraction by noting that both the Ly$\alpha$ and H$\alpha$ fluxes are related to the star formation rate of the galaxy. By comparing the Ly$\alpha$ SFR versus an independently measured SFR, 
we can calculate a crude estimate of the Ly$\alpha$ escape fraction (f$_{esc}$). f$_{esc}$ = SFR$_{Ly\alpha}$/SFR$_{BC03}$, where SFR$_{BC03}$ is the SFR predicted from Bruzual \& Charlot (2003) models.
 A similar technique was used in \citet{onoA} to measure the escape fractions of narrow-band LAEs at  $z=3-4$.

Using the spectroscopic Ly$\alpha$ redshifts, the {\it Le Phare} \footnote{$http://www.oamp.fr/people/arnouts/LE_PHARE.html$} SED fitting code was used to generate estimates of SFR, E(B-V) and stellar mass for the sources.
The SED fitting was performed following  \citet{ilbert2010}, with the redshifts of the model SEDs fixed to the spectroscopic redshifts of our sources.
Briefly, a set of  galaxy templates  were generated using Bruzual \& Charlot (2003) with exponentially declining SFRs, two metallicities, Calzetti et al. (2000) extinction,   and including emission features (Ly$\alpha$, [OII], [OIII], H$\beta$ and H$\alpha$). 
See Table 1 from  \citet{ilbert2010} for a list of the parameter values used.
Using a $\chi^2$ procedure,the templates were fit  to the multi-band  optical/near-infrared photometry taken from 
6 broad bands from the SuprimeCam/Subaru camera ($B_J$, $V_J$, $g^+$, $r^+$, $i^+$, $z^+$), 1 broad band 
from MEGACAM at CFHT ($u^{\prime}$), 14 medium and narrow bands 
from SuprimeCam/Subaru (IA427, IA464, IA484, IA505, IA527, IA574, IA624, IA679, IA709, IA738, IA767, IA827, NB711, NB816),  the $Y$,$J$,$H$, and $K_s$ broad bands from the Ultra-Vista survey of COSMOS (McCracken et al. 2012)\footnote{The Ultra-Vista data cover the central 1x1.5 degree area of the COSMOS survey in $Y$,$J$,$H$, and $K_s$ bands with an exposure time of 11.8, 13.8, 11.8, and 10.9h respectively.  The estimated $5\sigma$ depths are $Y=24.6$, $J=24.7$, $H=23.9$, $K_s=23.7$ AB.  Deeper IRAC data from several small programs targeting our spectroscopic area and the SEDS survey have also been included in the photometry, significantly improving the mass estimates for fainter targets.  These data reach an exposure time of 2-12h per pixel in the 3.6$\mu$m and 4.5$\mu$m bands.}
(in the region outside the survey coverage of the Ultra-Vista data the J-band from the WFCAM/UKIRT 
camera, H- and K-band from the WIRCAM/CFHT camera are used), 
and the 4 IRAC/{\it Spitzer} channels. From the fits, the median SFRs, and stellar masses are used along with the 16 and 84 percentile values are taken as the errors on for the SFR and stellar mass estimates. 
The errors on the SFRs and stellar masses are typically large (about an order of magnitude). The large uncertainties are due mostly to the faintness of the sources, since they are mostly detected at the 3-7$\sigma$ level in the photometry.
The E(B-V) value used is from best fit SED.  The results of the SED fitting are listed in table 4.

For 153 of the 244 sources with 3$\sigma$ Ly$\alpha$, the SED fitting produced a best fit SED with $\chi^2 < 50$ ($4$ $B_J$ LBGs, $16$ $g^+$ LBGs, $20$ $V_J$ LBGs,  $16$ $r^+$ LBGs, $2$ $i^+$ LBG,
$16$ IA624, $19$ NB711, $33$ NB816 sources, and $27$ from the various other selection methods), and the following analysis is restricted to these. The $\chi^2 < 50$ criteria was chosen after inspection of the best fit SED and photometric data points of each source.
For sources with $\chi^2 > 50$, the best fit SED was a bad match for 3 or more of the rest-frame UV and optical  data points. These sources may have properties outside of the parameter space covered by the galaxy models and hence the SED fitting may 
produce unreliable estimates, and so these sources were excluded from the subsequent analysis. For sources with $10< \chi^2 < 50$, these were the result of 1 to 2 discrepant photometric data points where the best fit SED matched the other data points within the errors.
We use the SFR values to estimate our escape fractions.
To convert our Ly$\alpha$ fluxes into SFRs, we first assume CASE-B recombination and convert the measured Ly$\alpha$ luminosities  into expected H$\alpha$ 
luminosities (L$_{H\alpha}$ = L$_{Ly\alpha}/8.7$) and then to SFRs using equation 2 in \citet{kennicutt}.
We plot the Ly$\alpha$ luminosity versus stellar mass and SFR in figures 8 and 9 respectively. No trend between the Ly$\alpha$ luminosity and either mass or SFR is observed. 
The LAEs tend to have higher Ly$\alpha$ luminosities than the LBGs, but he LBGs, NB711 and NB816 LAEs have similar stellar mass ($\sim10^{10}$~M$_{\odot}$), and SFRs ($\sim50$~M$_{\odot}$/yr).  
The IA624 LAEs on average have slightly lower stellar masses ($\sim5\times10^{9}$~M$_{\odot}$) and SFRs $\sim15$~M$_{\odot}$/yr) as these sources are on average 1 magnitude fainter in the rest-frame UV/optical.
Previously Yuma et al. (2010) compared the properties of 3 LAEs and 88 LBGs at $z\sim5$ and found that the physical properties of LAEs and LBGs occupy similar parameter spaces. At the same rest-frame UV or optical luminosity,
they found no difference in stellar properties (stellar mass, SFR, dust extinction) between their LAEs and LBGs at $z\sim5$.

In figure 10 we show  f$_{esc}$ versus redshift. A definite difference is seen between the escape fractions of narrow-band LAEs and the LBGs at fixed redshift, as the intermediate/narrow band sources have higher mean f$_{esc}$ and larger range of f$_{esc}$.
Yet there is essentially no change in the escape fraction for the LBG sources with redshift, nor is there a noticeable difference between the escape fractions of the NB711 and NB816 selected LAEs.  
The mean, median and range of f$_{esc}$ for each of the sub-samples
is listed in table 5. Our measured escape fractions for the NB816 sources in COSMOS have the same range of escape fractions  as the NB816 selected sources studied by \citet{onoB} in the Subaru/XMM-Newton Deep Survey field. Our mean 
escape fraction of 0.37  agrees with their value of 0.36. Our mean and median values are also in agreement with the escape fraction of $z~2.2$ LAEs 
studied by Hayes et al. (2010), who found median escape fraction to be  higher than 0.32.

In figure 11 we show changes in f$_{esc}$ with the stellar mass and E(B-V).  
There is a slight trend with decreasing escape fraction and increasing stellar mass. This is likely due to the trend for
more massive and luminous galaxies at higher redshifts to have higher dust extinctions \citep{bouwens}. Plotted versus E(B-V), we see an interesting trend where the sources with the highest extinctions have  low escape fractions (f$_{esc}\sim0.1$), 
but  sources with low extinctions  have a range of escape fractions. As extinction increases the range of the escape fraction decreases. This is similar to the trend seen for Ly$\alpha$ sources at $z\sim0.1$ (\citet{scarlata}, \citet{atek2}) and $z\sim3$ (Blanc et al. 2011).
This may indicate that the same physical conditions/processes (such as gas kinematics,  HI covering fraction, and/or galaxy morphology) that inhibit and allow for the escape of Ly$\alpha$ photons at low redshift are similarly occurring in high redshift galaxies too.

In order for this explanation to hold, sources lacking Ly$\alpha$ should be on average more dusty than sources without.
For $15$ spectroscopic sources with redshifts measured from absorption features, the Ly$\alpha$ 1$\sigma$ flux upper limits were calculated (see Table 6).
Using these upper limits and the SED SFRs for these sources, the upper limits for the escape fraction for these sources was also determined.
The combined escape fraction upper limit for these  sources is $0.8$\%. As expected these sources are offset from the Ly$\alpha$ sample with significantly higher $<$E(B-V)$> = 0.19$ than the mean for sources with Ly$\alpha$ detections.
Interestingly these sources have a slightly higher mean stellar mass $<$M$_*$$> = 2\times10^{10}$M$_{\odot}$ and have $<$SFR$> = 169$M$_{\odot}$/yr similar to the $V_J$ LBGs.

\section{Conclusion}	
In this paper we present an analysis of  a spectroscopic sample of $244$ LBGs and LAEs at $4<z<6$ in COSMOS with clear Ly$\alpha$ detections. 
We have attempted to determine variations in the Ly$\alpha$ properties for these sources and their evolution with redshift.
 The sources were targeted for spectroscopy using a range of high redshift selection techniques, including LBG, intermediate/narrow-band, photo-z, and IRAC CH2 detections.  
 The goal of the spectroscopic program was to select as complete a sample at $z>4$ as possible, for objects brighter than $z^+<25$ and more massive than $10^{10.5}M_\sun$ (Capak et al. in prep).
 We  measured EW$_{Ly\alpha,0}$ and escape fractions for $B_J$,$g^+$,$V_J$,$r^+$,$i^+$ LBGs, one intermediate-band and two narrow-band selected samples of LAEs at $z\sim4.2$, $z\sim4.8$, and $z\sim5.6$.
 A sub-sample of $153$ sources have estimates of E(B-V), SFR and M$_\sun$ from SED modeling. We analyze the variations of the Ly$\alpha$ properties for this subset with respect to these parameterizations of the host galaxies.
The results are summarized below.

1) We find that the Ly$\alpha$ EWs remain roughly constant with redshift for both the LBG  and intermediate/narrow-band LAEs. 
While low EW$_{Ly\alpha,0}$ are detected for sources at all redshifts, increasingly larger EW$_{Ly\alpha,0}$ are measured for sources from samples at higher redshifts. These results are in accordance with the results of Stark et al. 2010 who found a similar trend
for LBGs with Ly$\alpha$ at $z=3-6$, and with the similar findings of \citet{nilsson} studying LAEs at lower redshifts (z=$2-3$). The speculation is that the change in EW distributions with redshift is the result of increased dust content  in LAEs at lower redshifts, but
this is yet to be confirmed.

2) No trends were found between Ly$\alpha$ luminosity  and stellar mass or SFR. Except for the IA624 LAEs, which on average have lower UV luminosities, the sources tend to have similar stellar masses and SFRs. 
The mean Ly$\alpha$ luminosities are slightly higher for the LAEs than the LBGs.

3) We find that the Ly$\alpha$ escape fraction of narrow-band LAEs is, on average, 
higher and has a larger variation than LBG selected sources. The escape fraction does not show a dependence on redshift.  Our escape fraction for NB816 LAEs, 0.48, agrees within the errors to 
 escape fraction of  NB816 selected sources measured by \citet{onoB} in the Subaru/XMM-Newton Deep Survey field (0.36), and the mean escape fraction of  Ly$\alpha$ sources (0.32) at $z=2.2$ studied by  Hayes et al. (2010).

4) Similar to what has been found for sources with Ly$\alpha$ emission at low redshifts, the sources with the highest extinctions show the lowest escape fractions. The range of escape fractions increases with decreasing extinction.
This is evidence that the dust extinction is the most important factor affecting the escape of Ly$\alpha$ photons, but at low extinctions other factors such as 
HI covering fraction and gas kinematics can be just as effective at inhibiting the escape of Ly$\alpha$ photons.

\acknowledgements
Based in part on data obtained at the W. M. Keck Observatory, which is operated as a scientific partnership among the California Institute of Technology, the University of California, and NASA 
and was made possible by the generous financial support of the W. M. Keck Foundation.

\clearpage


\begin{figure}
\plotone{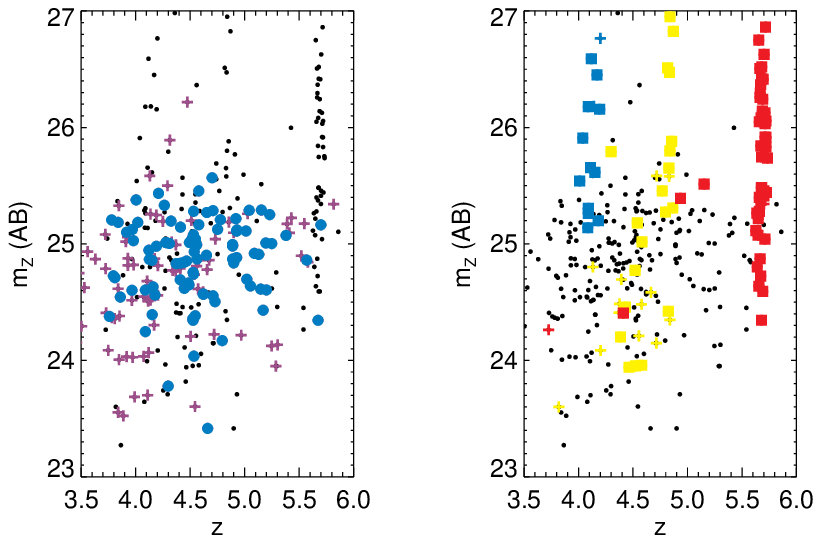}
\caption{Redshift versus apparent z (AB) magnitude.In both panels the black dots represent all sources with measured spectroscopic redshifts. 
{\it Left Panel}: The m$_z$-redshift distribution for LBG selected sources with (without) Ly$\alpha$ as blue circles (purple crosses).
{\it Right Panel}: The m$_z$-redshift distribution for narrow-band selected sources with (without) Ly$\alpha$ detections.
Blue squares represent the IA624 sources, the yellow squares (crosses) the NB711 sources, and the red squares (crosses) the NB816 sources. 
The four low-z NB816 outliers are from the relaxed color-cut criteria used to select the LAEs  at $z\sim5.6$, and would not have made the more stringent cut from Murayama et al. (2007).
For both the LBG and intermediate/narrow band selected sources, Ly$\alpha$ detection shows no bias by either redshift,  or magnitude, and hence 
luminosity, with regards to Ly$\alpha$ detection down to the detection limits of the spectroscopy. However, the narrow-band sources with Ly$\alpha$ are on average 0.8 magnitudes fainter than the LBG sources with Ly$\alpha$}
\end{figure}
\clearpage

\begin{figure}
\epsscale{0.85}
\plotone{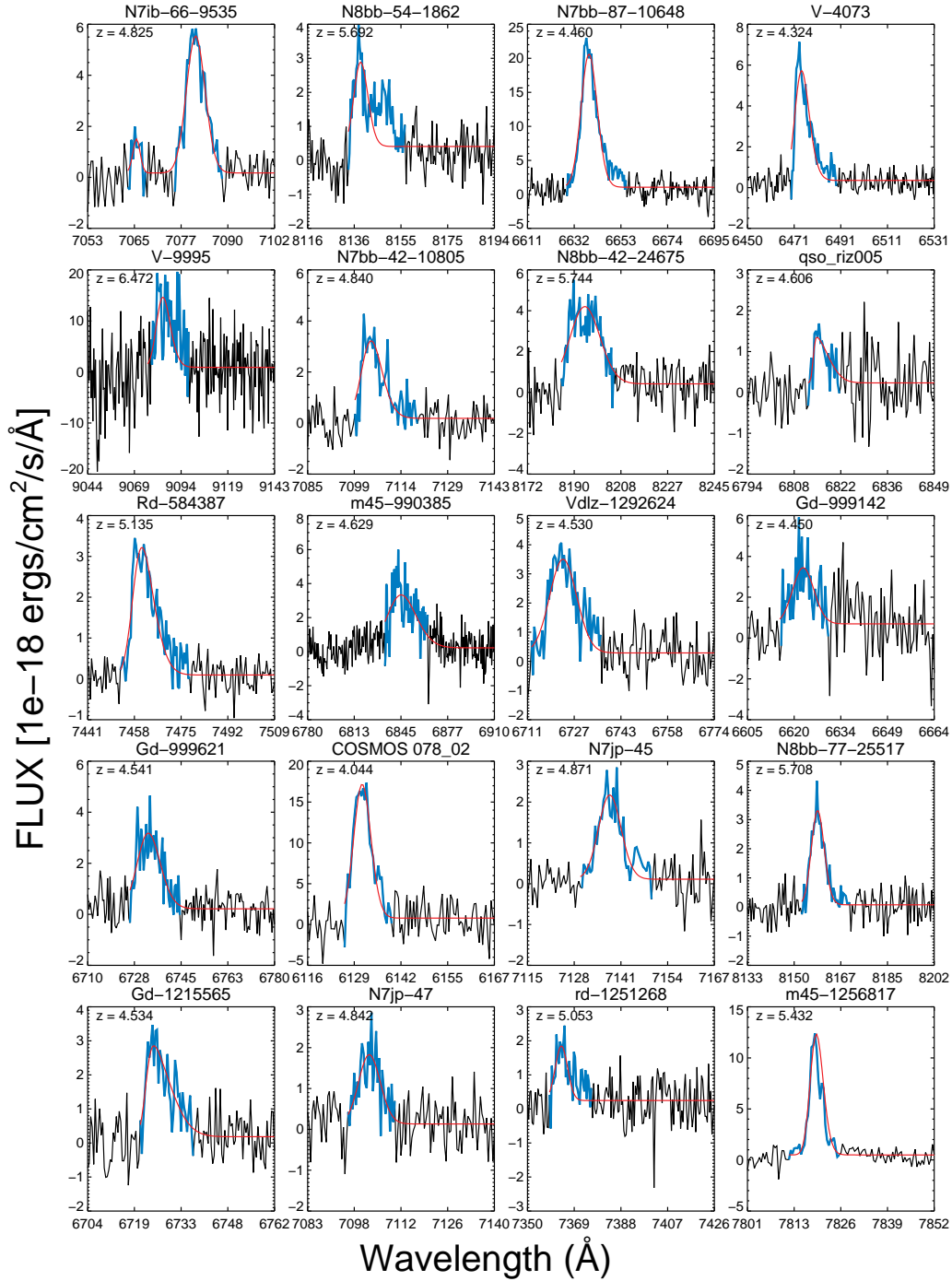}
\caption{Spectra of 20 sources randomly chosen, showing the Ly$\alpha$ emission feature. The blue line highlights the region of each spectrum used for the numerical integration. The red line shows the best skewed Gaussian fit to the data.
The 1D and 2D spectra will be shown in the data paper (Capak et al. in prep).}
\end{figure}
\clearpage

\begin{figure}
\plotone{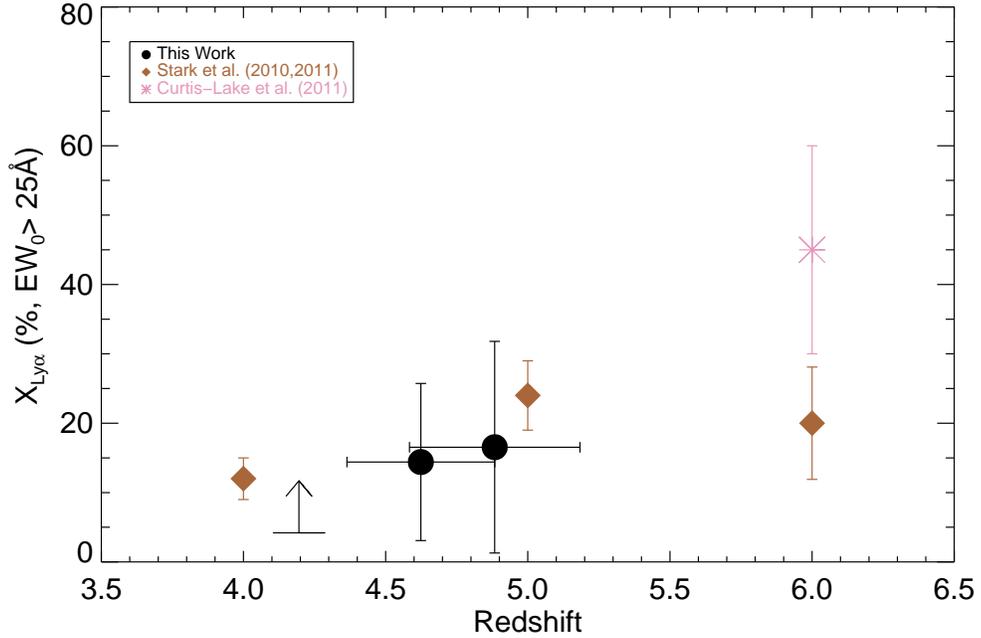}
\caption{ The fraction of LBGs with EW$_{Ly\alpha,0} > 25$ and  $-21.75 <$~M$_{UV} < -20.25$ plotted versus mean redshift.
Plotted is the fraction of  $B_J$ + $g^+$ LBGs (lower limit) at $z\sim4.2$,   $V_J$ LBGs (filled circle) at $z\sim4.6$, and $r^+$ LBGs (filled circle) at $z\sim5.0$.
Other fractions are taken from Curtis Lake et al. (2011) and Stark et al. (2010, 2011). Our measured fractions do not point to an evolution of the 
Ly$\alpha$ fraction of luminous LBGs over the redshift range $3.8 < z < 5.5$ but are consistent with the fractions reported in Stark et al. (2010, 2011).
}
\end{figure}
\clearpage

\begin{figure}
\plotone{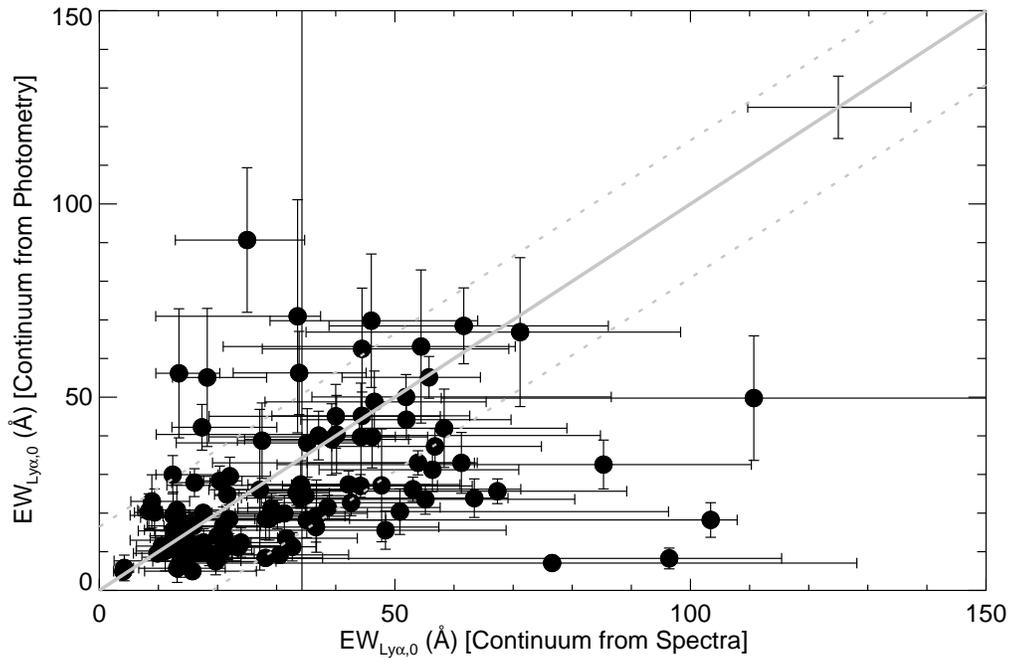}
\caption{The flux calibrated rest frame Ly$\alpha$ equivalent width comparison between continuums measured using the spectra, and continuums measured using the photometry.
The solid gray line shows a  1 to 1 correspondence, and the dashed gray lines show  the $1\sigma$ deviation from a 1 to 1 correspondence determined from the mean errors on both equivalent widths.
The mean equivalent width error bar is plotted in the upper right corner.
}
\end{figure}
\clearpage

\begin{figure}
\plotone{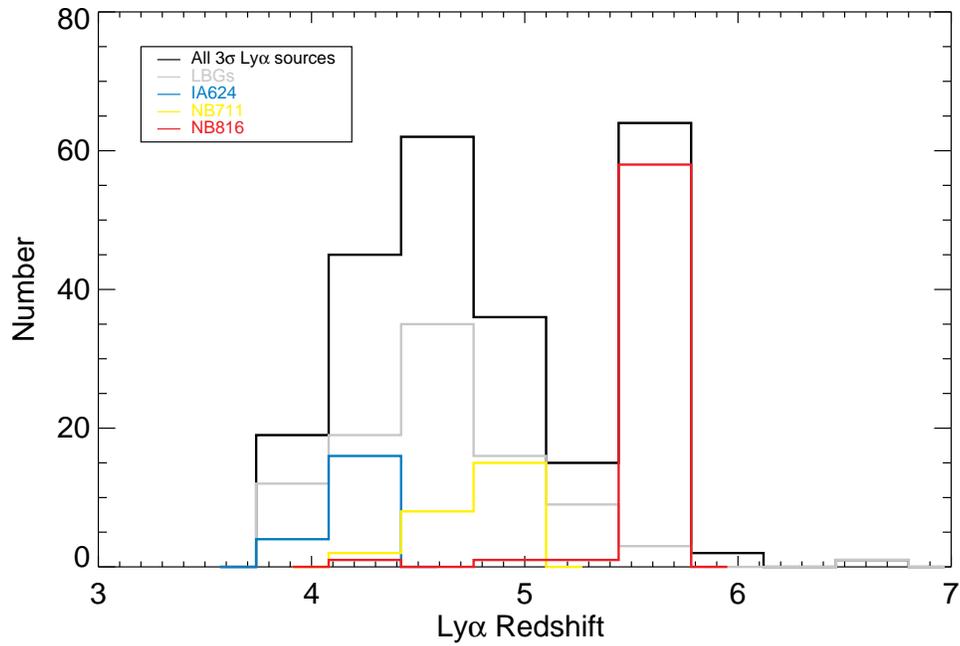}
\caption{Redshift distribution of the Ly$\alpha$ sample. Sources are divided into the following categories: All sources (black), LBGs (gray), IA624 (blue), NB711 (yellow), NB816 (red).
The source selection for each of these sub-samples is described in section 2.2}
\end{figure}
\clearpage

\begin{figure}
\plotone{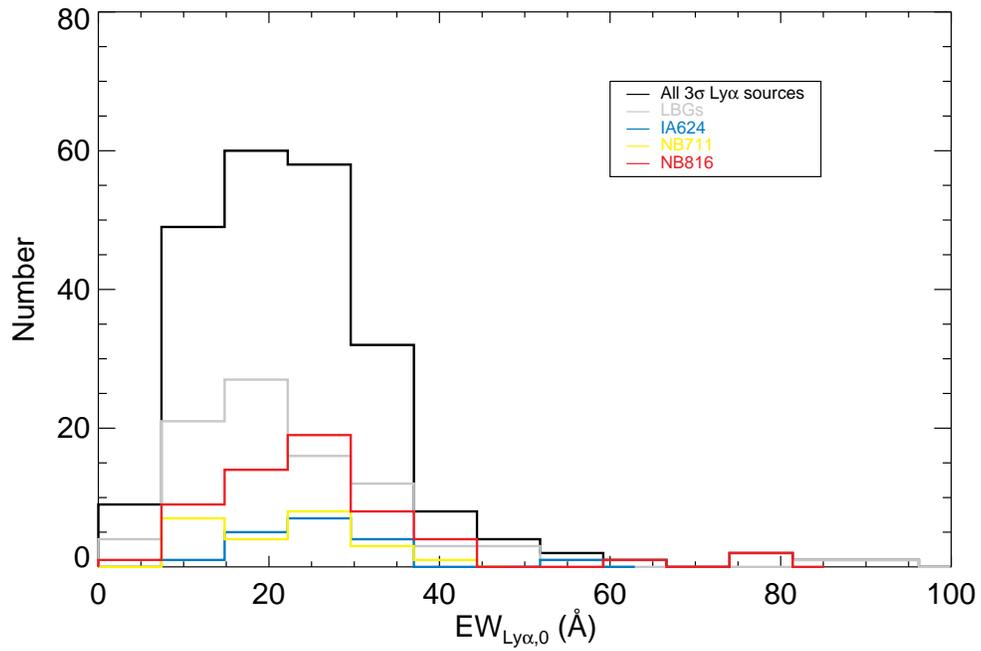}
\caption{The flux calibrated rest frame Ly$\alpha$ equivalent width distribution. Sources are divided into the following categories: All sources (black), LBGs (gray), IA624 (blue), NB711 (yellow), NB816 (red). 
The LBGs  have a lower mean EW than the narrow-band LAEs, which may be due to the narrow-band LAEs being on average fainter than the LBGs by 0.8 magnitudes.}
\end{figure}
\clearpage

\begin{figure}
\plotone{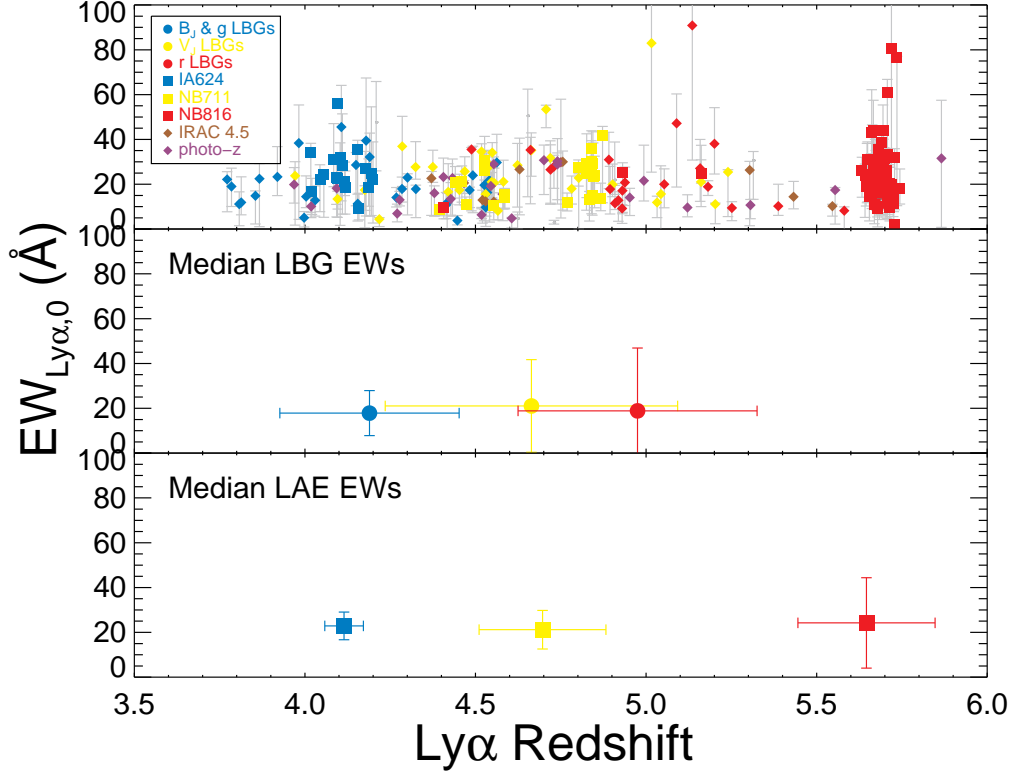}
\caption{The change in rest frame Ly$\alpha$ equivalent width  as a function of redshift. The median EW$_{Ly\alpha,0}$ of both the LBGs and LAEs show no evolution with redshift. The LAEs tend to have slightly
higher EWs than the LBGs at similar redshifts.
{\it Top Panel}: EW$_{Ly\alpha,0}$ versus redshift for the entire sample. {\it Middle Panel}: The median values  of EW$_{Ly\alpha,0}$ and redshift for each of the LBG sub-samples. The median EW$_{Ly\alpha,0}$ 
shows no evolution with redshift for the LBG selected sources, though the sample variance increases with redshift.
{\it Bottom Panel}: The median values  of EW$_{Ly\alpha,0}$ versus redshift for each of the intermediate/narrow band LAEs. Similar to the LBGs, the median EW$_{Ly\alpha,0}$ 
shows no evolution with redshift.  The EW, redshift error bars are the sample variances.
The filled circles represent the LBG sources, and are colored as follows: 
The blue-dots represent  $B_J$ and $g^+$ LBGs, yellow-dots the $V_J$ LBGs, red dots the $r^+$ LBGs and violet dots the $i^+$ LBGs. The filled squares represent the narrow-band selected 
LAEs with the blue-squares for the $z\sim4.2$ sources, the yellow-squares for the NB711 sources and the red-squares for the NB816 sources. The brown-diamonds represent the 
other selected sources. }
\end{figure}
\clearpage

\begin{figure}
\plotone{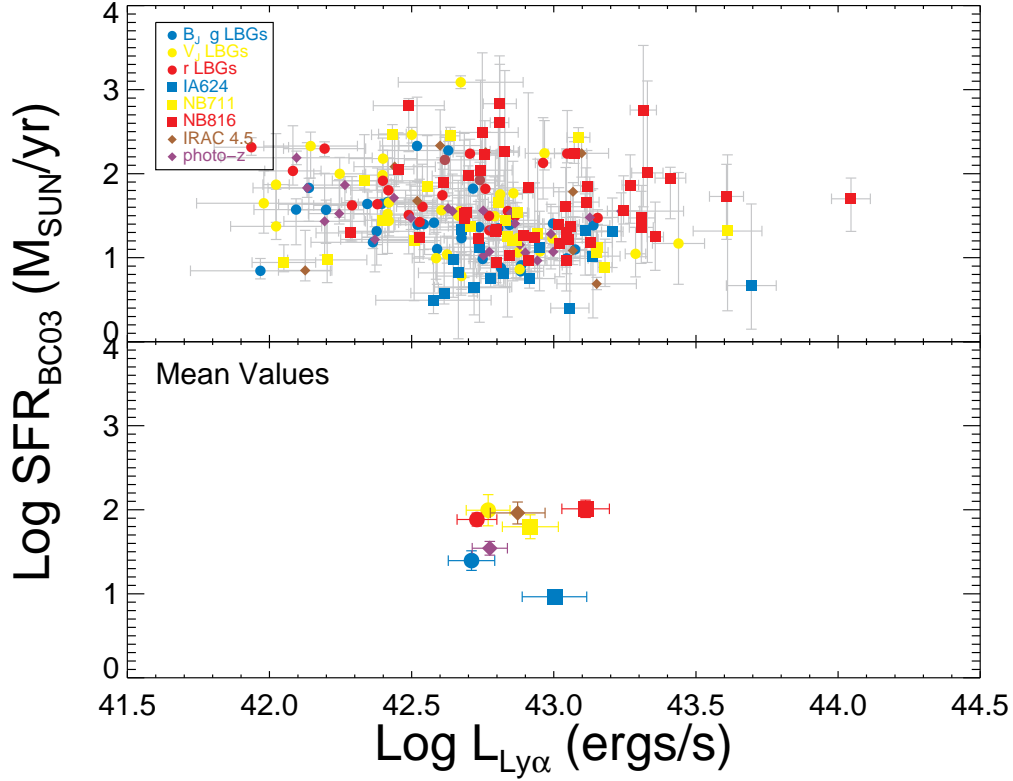}
\caption{ The flux calibrated Ly$\alpha$ luminosity plotted versus SFR estimated from BC03 galaxy models. {\it Top Panel}: All $153$ sources with measured SFRs. {\it Bottom Panel}: The mean and error on the mean
of the Ly$\alpha$ luminosity and SFR  for each of the sub-samples. 
No particularly strong trends are found between Ly$\alpha$ luminosity, and SFR. 
The LAEs on average have higher Ly$\alpha$ luminosities. 
All have similar distributions of SFR except for the IA624 sources, which  have $\sim1$~magnitude fainter UV luminosities than the rest of the LBGs and LAEs, and slightly lower SFRs. 
 The symbols are the same as in figure 7. }
\end{figure}
\clearpage

\begin{figure}
\plotone{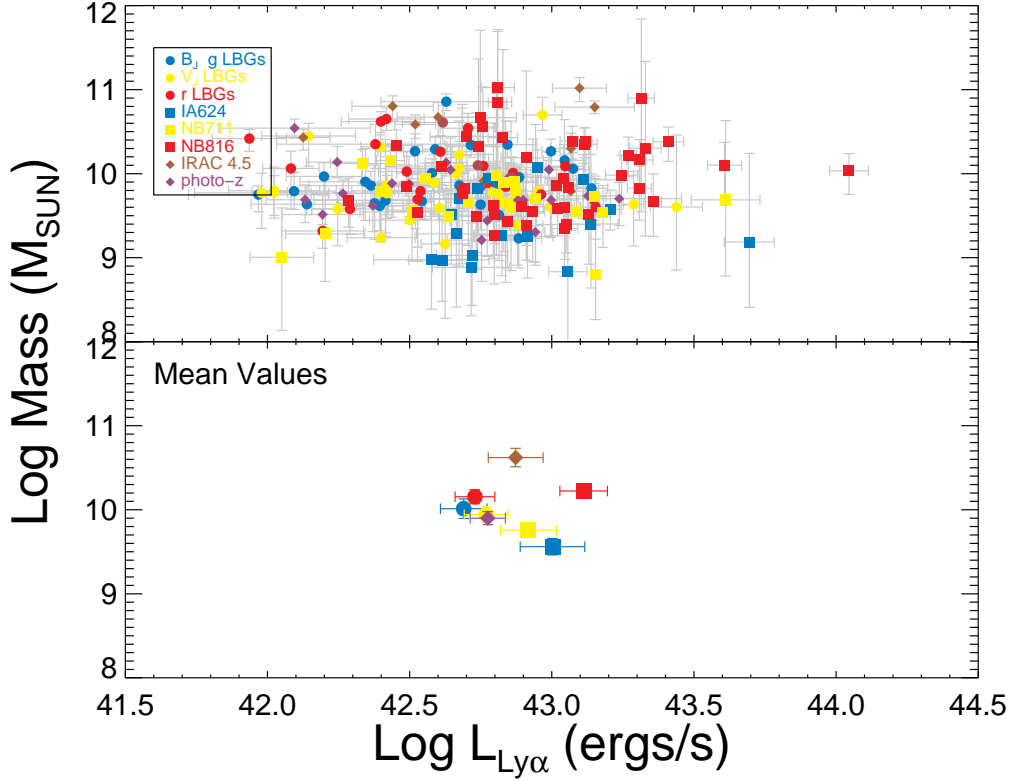}
\caption{ The flux calibrated Ly$\alpha$ luminosity plotted versus stellar mass estimated from BC03 galaxy models in \citet{ilbert2010}.  {\it Top Panel}: All $153$ sources with measured stellar masses. {\it Bottom Panel}: The mean and error on the mean
of the Ly$\alpha$ luminosity and stellar mass  for each of the sub-samples. 
 Similar to the figure 7, no particularly strong trends are found between Ly$\alpha$ luminosity and stellar mass. The LBGs and LAEs all have very similar distributions of stellar mass, except the 
 IA624 sources which are slightly less massive.
 The symbols are the same as in figure 7. }
\end{figure}
\clearpage

\begin{figure}
\plotone{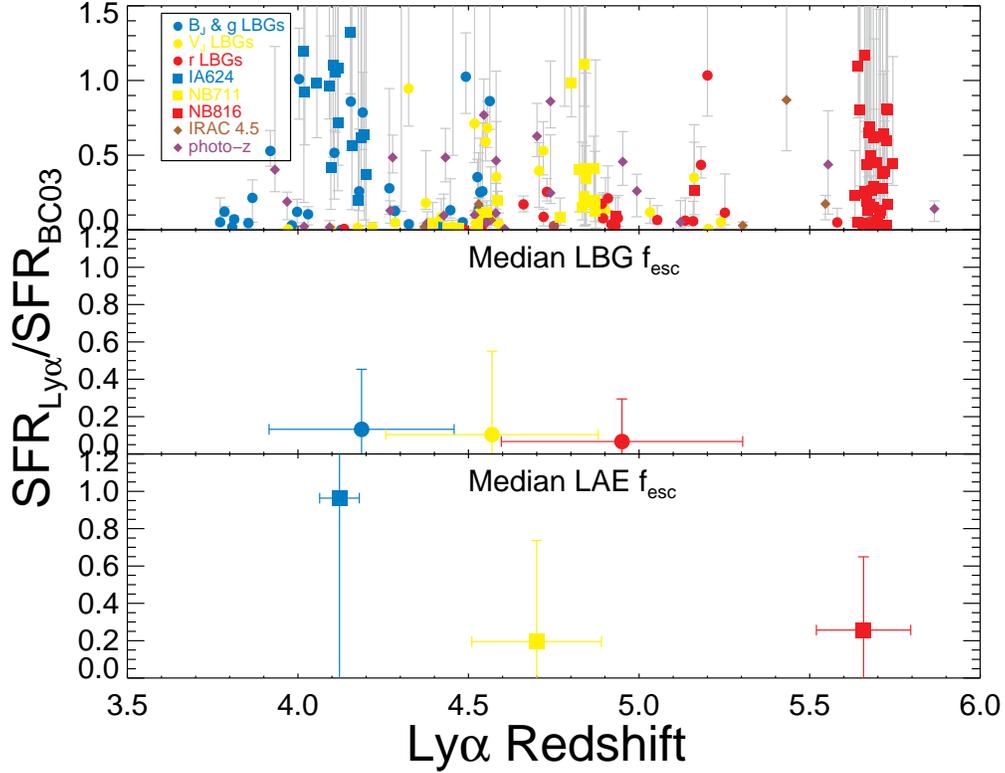}
\caption{Estimated Ly$\alpha$ escape fraction plotted versus redshift. The symbols are the same as in figure 6. {\it Top Panel}: All 153 sources with SFRs from SED fitting. {\it Middle Panel}: The median escape fractions of the LBGs, with the
error bars showing the sample variances. {\it Bottom Panel}: The median escape fractions of the LAEs, with the error bars showing the sample variances. 
The majority of sources indicate escape fractions at or below $50$\%. The escape fractions are highly uncertain
due to uncertainties in the SED SFRs. The LAEs have the largest uncertainties due to the faintness of theses sources which results in larger photometric  errors and greater uncertainties in the physical properties derived from
the SED fits.
The sources with the highest escape fractions are narrow/intermediate band selected LAEs. The median escape fraction for the entire sample is $18$\%. The data is consistent with no change in escape fraction with redshift for the LBGs.
The NB711 and NB816 LAEs have similar mean and median escape fraction twice that of the LBGs. The IA624 sources have extremely high escape fractions, with mean and median values  up to and exceeding f$_{esc}\sim1$. The high values  are likely attributable to  the uncertainties of the SED derived SFRs as these source were chosen to be  faint, m$_z > 25$(AB). The top panel shows the entire sample, the middle panel shows the median values for the LBGs, and the bottom panel shows the median values for the narrow band LAEs.
}
\end{figure}
\clearpage

\begin{figure}
\plotone{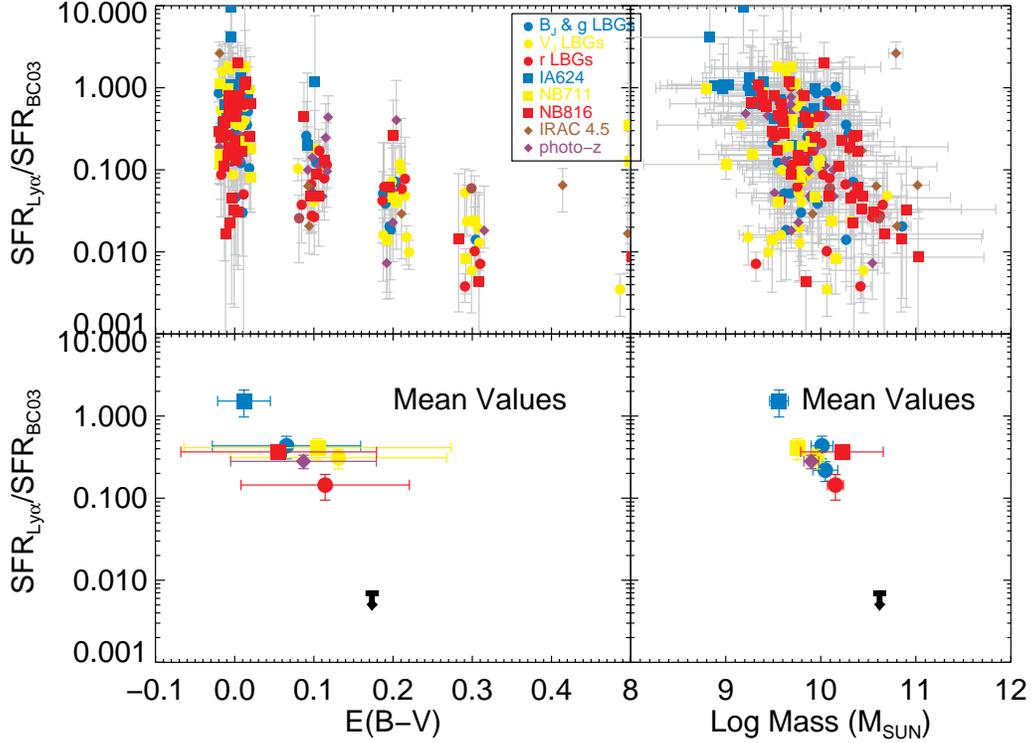}
\caption{ {\it Left}: Estimated Ly$\alpha$ escape fraction plotted versus extinction estimated from BC03 models.  The E(B-V) values are discrete at 0.0, 0.1, 0.2, 0.3, 0.4, and 0.5. 
To make the points more visible a random scatter of 0.02 has been added to their values. 
This shows that while extinction inhibits the escape of Ly$\alpha$ photons, there are other factors that govern Ly$\alpha$ escape such as  HI covering fraction, and gas kinematics, that can inhibit its escape even when there is little dust.
{\it Right}: The  Ly$\alpha$ escape fraction is plotted versus stellar mass estimated from BC03 models. There is a slight trend between stellar mass and escape fraction, 
with higher stellar mass sources having lower escape fractions.
The black arrows represent  the combined upper limit on the escape fraction for $15$ spectroscopic sources with only  1$\sigma$  Ly$\alpha$ flux upper limits. 
The E(B-V) and M$_*$ values plotted are the mean values for these sources.
The other symbols are the same as in figure 7. 
The panels on the lower left and right  show the mean values for each of the source types.
The error bars on the mean E(B-V) values represents the sample variance, while the mean f$_{esc}$ and M$_*$ error bars are the errors on the means.
}
\end{figure}
\clearpage

\begin{deluxetable}{rrrrr}
\tabletypesize{\scriptsize}
\tablecaption{DEIMOS Sources with Ly$\alpha$ Emission}
\tablewidth{0pt}
\tablecolumns{5}
\tablehead{ \colhead{Type}  &\colhead{\# $>3\sigma$ Ly$\alpha$}  &\colhead{\# with EW$_{Ly\alpha,0} > 25$\AA} &\colhead{AGN with $>3\sigma$ Ly$\alpha$ } &\colhead{\# Observed} 
  }
\startdata
All LBGs:&                          95&      32&     1&    380\\
$B_J$ LBGs:&                   10&        3&       0&  49\\
$g^+$ LBGs:&                   21&        3&     0& 158\\
$V _J$ LBGs:&                    39&       16&    1&    101\\
$r^+$ LBGs:&                     23&         9&    0&  56\\
$i^+$ LBGs:&                      2&          1&     0&   16\\
IA624:&                            20&         9&     2&     26\\
NB711:&                          25&          9&     1&   42\\
NB816:&                          61&          26&    0&   73\\
IRAC4.5:&                       11&            3&   1&    55\\
Photo-z:&                         22&           5&   0&     58\\
Other:&                               5&            0&    1&    10\\       
Total:&                            244&         84&   7&  644\\
\enddata
\end{deluxetable}

%

\begin{deluxetable}{rrrrrrrrr}
\tabletypesize{\scriptsize}
\tablecaption{Ly$\alpha$ Emission}
\tablewidth{0pt}
\tablecolumns{9}
\tablehead{\colhead{Source}& \colhead{RA J2000}& \colhead{DEC J2000}& \colhead{Type}& \colhead{z}& \colhead{Flux}& \colhead{EW$_{Ly\alpha,0}$} &\colhead{FWHM} &\colhead{Skew} \\
\colhead{ }& \colhead{ }& \colhead{ }& \colhead{ }& \colhead{ }& \colhead{(1e-18  ergs/cm$^2$/s)}& \colhead{(\AA)}& \colhead{(\AA)}& \colhead{ }
  }
\startdata
           N7ib-66-9535  &149.967958&  2.258167&        NB711&  4.825&      30.1$\pm$ 6.00&      24.9$^{+6.11}_{-16.48}$&      5.72$\pm$ 1.47&       0.80$\pm$ 0.01\\
           N8bb-54-1862  &149.971875&  2.118167&     NB816& 5.692&      19.8$\pm$ 7.60&       5.3$^{+2.63}_{- 4.10}$&      7.23$\pm$ 1.79&           1.12$\pm$  0.10\\        
           N8bb-54-20446 &149.933583&  2.014083&    NB816& 5.688&      15.5$\pm$ 2.62&      27.8$^{+11.65}_{-13.58}$&      9.01$\pm$ 2.22&      1.51$\pm$      0.05\\
           N8bb-66-30821 &149.942250&  2.128583&     NB816&5.666&      18.1$\pm$ 2.18&      24.9$^{+4.54}_{- 1.78}$&      9.76$\pm$ 2.34&          2.34$\pm$      0.10\\
           N8jp-66-40    &149.977208&  2.254611&         NB816& 5.688&      13.8$\pm$ 2.10&      20.0$^{+7.44}_{-13.25}$&      6.80$\pm$ 1.89&         0.92$\pm$      0.07\\
           N8jp-66-41    &149.978292&  2.177611&         NB816&  5.662&      31.7$\pm$ 4.83&      15.9$^{+6.03}_{- 6.76}$&      6.26$\pm$ 1.56z&        0.52$\pm$      0.07\\
           B-8431        &149.941292&  2.057139&            $B_J$ LBG& 4.150&      31.6$\pm$ 8.99&      28.6$^{+5.55}_{-18.22}$&      9.13$\pm$ 2.26&      2.31$\pm$      0.10\\
           N8bb-37-10756 &150.790833&  1.897889&   NB816&  5.705&      46.6$\pm$ 5.45&      21.4$^{+2.98}_{-14.10}$&      4.54$\pm$ 0.15&      0.98$\pm$     0.01\\
\enddata
\end{deluxetable}

\clearpage

\begin{deluxetable}{rrrrrrrr}
\tabletypesize{\scriptsize}
\tablecaption{Double Peaked Ly$\alpha$ Emission}
\tablewidth{0pt}
\tablecolumns{8}
\tablehead{\colhead{Source}& \colhead{RA J2000}& \colhead{DEC J2000}& \colhead{Type}& \colhead{Ly$\alpha$ z}& \colhead{Flux (1e-18  ergs/cm$^2$/s)}& \colhead{EW (\AA)} &\colhead{FWHM(\AA)}
  }
\startdata
pz-559631& 150.127833&   1.862111& photo-z&              4.262&       16.2$\pm$ 2.83&       160.9$^{+26.22}_{- 109.19}$&       4.89$\pm$ 1.20\\
                 ~&                     ~&                  ~&               ~&            4.278&       42.3$\pm$ 3.45&       68.8$^{+20.03}_{- 14.32}$&       4.28$\pm$ 2.22\\
m45-786441& 150.142917&   1.989222& IRAC CH2&      4.457&        7.8$\pm$ 0.74&        14.0$^{+ 5.19}_{- 3.53}$&         2.26$\pm$ 0.18\\
                ~&                       ~&                    ~&              ~&          4.466&       54.4$\pm$ 2.05&      112.6$^{+ 5.87}_{- 25.81}$&       6.77$\pm$ 2.33\\
\enddata
\end{deluxetable}

\clearpage

\begin{deluxetable}{rrrrrrrr}
\tabletypesize{\scriptsize}
\tablecaption{Best-fit Model SED Parameters}
\tablewidth{0pt}
\tablecolumns{8}
\tablehead{\colhead{Source}& \colhead{RA J2000} &\colhead{DEC J2000}& \colhead{Best $\chi^2$} &\colhead{Best E(B-V)}  &\colhead{Log Median SFR$^a$} &\colhead{Log Median Mass$^a$} &\colhead{Median Age$^a$} \\
\colhead{  }&  \colhead{ }&  \colhead{  }& \colhead{  }& \colhead{  }&  \colhead{(M$_{\odot}$/yr)}&  \colhead{(M$_{\odot}$)}&  \colhead{(Gyr)}
  }
  \startdata
N8jp-109-108  &150.805417&  2.925000&      21.2&       0.0&      1.31$^{ 2.22}_{ 0.89}$&      9.62$^{10.60}_{ 8.99}$&      0.23$^{ 0.58}_{ 0.09}$\\
N8bb-54-1862  &149.971875&  2.118167&      12.3&       0.2&      2.24$^{ 2.38}_{ 2.12}$&     10.39$^{10.56}_{10.18}$&      0.17$^{ 0.30}_{ 0.11}$\\
N8bb-54-20446 &149.933583&  2.014083&      11.4&       0.0&      1.23$^{ 1.67}_{ 0.80}$&      9.49$^{ 9.82}_{ 9.06}$&      0.22$^{ 0.55}_{ 0.10}$\\
N8bb-66-30821 &149.942250&  2.128583&     109.4&       0.0&      1.10$^{ 1.48}_{ 1.01}$&      9.97$^{10.27}_{ 9.67}$&      0.62$^{ 0.85}_{ 0.30}$\\
N8jp-66-40    &149.977208&  2.254611&       0.6&       0.0&      1.47$^{ 2.40}_{ 0.65}$&      9.76$^{10.78}_{ 8.93}$&      0.25$^{ 0.59}_{ 0.10}$\\
N8jp-66-41    &149.978292&  2.177611&       3.6&       0.0&      1.61$^{ 2.07}_{ 0.89}$&      9.94$^{10.35}_{ 9.33}$&      0.27$^{ 0.61}_{ 0.11}$\\
B-8431        &149.941292&  2.057139&     156.0&       0.0&      1.12$^{ 1.19}_{ 1.04}$&      9.80$^{ 9.96}_{ 9.57}$&      0.65$^{ 0.97}_{ 0.36}$\\
V-2019        &149.941750&  2.111778&       ---&       0.0&      0.00$^{ 0.00}_{ 0.00}$&      0.00$^{ 0.00}_{ 0.00}$&      0.00$^{ 0.00}_{ 0.00}$\\
\enddata
\tablenotetext{a}{The superscripts (subscripts) represent the 84\% (16\%) values of the likelihood distribution from the SED fitting}
\end{deluxetable}

\clearpage

\begin{deluxetable}{rccr}
\tabletypesize{\scriptsize}
\tablecaption{Ly$\alpha$ Escape Fractions}
\tablewidth{0pt}
\tablecolumns{4}
\tablehead{ \colhead{Type}& \colhead{Mean f$_{esc}$}& \colhead{Median $f_{esc}$}& \colhead{ $\sigma_{fesc}$} 
  }
\startdata
$B_J$ \& $g^+$ LBGs:&	0.29&            0.13&   0.32\\
$V_J$ LBGs:&                    0.30&            0.10&    0.45\\
$r^+$ LBGs:&                     0.14&            0.07&   0.22\\
IA624:&				      1.51&            0.96&   2.27\\
NB711:&                                    0.41&            0.20&   0.54\\
NB816:&                                    0.37&            0.26&   0.39\\
\enddata
\end{deluxetable}

\begin{deluxetable}{rrrrr}
\tabletypesize{\scriptsize}
\tablecaption{Ly$\alpha$ Emission 1$\sigma$ Upper Limits}
\tablewidth{0pt}
\tablecolumns{5}
\tablehead{\colhead{Source}& \colhead{RA J2000}& \colhead{DEC J2000}& \colhead{z}& \colhead{Flux Upper Limit} \\
\colhead{ }& \colhead{ }& \colhead{ }& \colhead{ }& \colhead{ (1e-18  ergs/cm$^2$/s)}   }
\startdata
               m45-845998&     149.653809&  2.084128&     4.080&       8.5\\
              Vdlz-528373&      150.248474&  1.896556&     4.540&       6.5\\
            N7bb-77-37461&      150.191086&  2.317983&     4.376&       6.4\\
             Vdlz-1475339&      149.898865&  2.566839&     4.504&      11.1\\
              m45-1492079&      149.869263&  2.617303&     4.274&      11.5\\
               pz-1232157&      150.225754&  2.387444&     4.276&       8.6\\
                Id-533224&      150.297577&  1.868394&     5.430&       4.8\\
                pz-561143&      150.156738&  1.851828&     3.885&       7.1\\
                id-122195&     150.035584&  1.934689&     5.580&       3.8\\
                id-110783&       150.235519&  1.888269&     5.410&       5.8\\
               Rc-27-8213&     150.404953&  1.751894&     4.969&       7.9\\
                   B-4667&          150.595856&  1.914678&     4.169&      13.8\\
            N7bb-88-31418& 150.390366&  2.510094&     4.203&       0.2\\
            N7bb-30-38883&149.938736&  1.657944&     4.372&       8.3\\
            N7bb-50-39856&150.680740&  1.989203&     4.578&       5.7\\

\enddata
\end{deluxetable}

\end{document}